\documentclass[journal,onecolumn,12pt]{IEEEtran}
\usepackage[T1]{fontenc}
\usepackage{graphicx}

\begin{document}

\title{When Robots Say No: \\Temporal Trust Recovery Through Explanation}

\author{Nicola Webb\inst{1}\orcidID{0000-0003-0503-8641} \and Zijun Huang\inst{1} \and \\
Sanja Milivojevic\inst{2}\orcidID{0000-0001-8533-4699}
 \and Chris Baber\inst{3}\orcidID{0000-0002-1830-2272} \and  Edmund R. Hunt\inst{1}\orcidID{0000-0002-9647-124X} 
}

\author{Nicola Webb\textsuperscript{1}\footnote{[1] School of Engineering Mathematics and Technology, University of Bristol, UK. \\ \indent \indent \indent email: nicola.webb@bristol.ac.uk, edmund.hunt@bristol.ac.uk},
Zijun Huang\textsuperscript{1}, Sanja Milivojevic\textsuperscript{2}\footnote{[2] Bristol Digital Futures Institute, University of Bristol, UK}, Chris Baber\textsuperscript{3}\footnote{[3] School of Computer Science, University of Birmingham, UK} and Edmund R. Hunt\textsuperscript{1}} 

\maketitle              

\begin{abstract}
Mobile robots with some degree of autonomy could deliver significant advantages in high-risk missions such as search and rescue and firefighting. Integrated into a human-robot team (HRT), robots could work effectively to help search hazardous buildings. User trust is a key enabler for HRT, but during a mission, trust can be damaged. With distributed situation awareness, such as when team members are working in different locations, users may be inclined to doubt a robot's integrity if it declines to immediately change its priorities on request. In this paper, we present the results of a computer-based study investigating on-mission trust dynamics in a high-stakes human-robot teaming scenario. Participants (n = 38) played an interactive firefighting game alongside a robot teammate, where a trust violation occurs owing to the robot declining to help the user immediately. We find that when the robot provides an explanation for declining to help, trust better recovers over time, albeit following an initial drop that is comparable to a baseline condition where an explanation for refusal is not provided. Our findings indicate that trust can vary significantly during a mission, notably when robots do not immediately respond to user requests, but that this trust violation can be largely ameliorated over time if adequate explanation is provided. 
\end{abstract}

\section{Introduction}

Robotic systems have the potential to provide significant assistance in high-stakes missions such as search and rescue or firefighting, as they could reduce the personal safety risks to users and enhance effectiveness \cite{delmerico2019}. In the future, robots could obtain -- and be permitted -- a higher level of autonomy than the teleoperated systems in use today, for instance in being able to adjust their task priorities in response to dynamic environments. A human-robot team (HRT) deployed on a firefighting mission is likely to become physically distributed across the space, with agents in different locations as they search for people to rescue and fires to put out, for instance. Each agent will have a partial view of the mission context: their local perceptions, inference of what goal they should be pursuing, and expectations of what goals their teammates should be pursuing. Therefore, it is likely that the team's situation awareness will be distributed \cite{stanton2006}. Trust between teammates, a critical enabler of their cooperation (e.g. \cite{huang2021}), is expected to vary according to the developing situation faced by each teammate \cite{hunt2023}. 

We envisage situations occurring when human users may give commands to an autonomous robot teammate, but because the robot has revised its priorities -- for instance, it is urgently assisting a third party such as a bystander -- the robot may legitimately decline to change its task immediately. In this case, given the user's limited situation awareness, trust may be violated, for instance because the user may question the robot's integrity (alignment with team goals). This is in addition to trust violations that may occur for other reasons, such as robot mistakes or failures that would undermine perceptions of capability. Thus, we see that user trust will need to be anticipated and managed by the robot system \cite{hunt2023,webb2024co}, and may require explanations or other trust repair strategies following such a trust violation event \cite{baker2018toward}.  This is perhaps especially necessary in high stakes situations, where teammates have to rely on each other for assistance. In this work, we therefore sought to investigate two main research questions: first, how does trust between a human and robot dynamically evolve over time, when a robot declines a command; and second, how does providing an explanation of the robot's refusal improve trust recovery, compared to when no such information is given. Given the difficulty of conducting research with human participants in real hazardous environments, we conduct a computer-based study. 

In Section 2 we provide a brief overview of some work related to our research questions, and in Section 3 we describe our methodology. Section 4 presents results, including trust scores over time for the two conditions (baseline refusal versus explanation). Section 5 presents discussion and brief ideas for future work, before a conclusion in Section 6. 

\section{Related Work}
\subsection{Trust in human-robot teaming}

We have previously presented some of our thinking about trust in HRT in \cite{hunt2023} and results from a real-world experiment in \cite{milivojevic2024swift}. In this experiment participants had to work alongside two rover robots to search an unfamiliar environment, much as a firefighter might have to. Participants had not worked with the robots before, and yet such ad-hoc teams are not uncommon in emergency scenarios such as search and rescue \cite{ribeiro2021helping}. In ad-hoc teams, humans rely on a provisional level of `swift trust', which necessitates quickly rebuilding trust after any disruptions or conflicts. This quick trust restoration is vital for maintaining momentum and ensuring the success of a mission, especially when team members may not have prior experience working together. To continue working on missions successfully, it is important to implement trust-repair strategies that can swiftly address issues and restore confidence. Although such strategies often serve as short-term solutions, they can be sufficient to achieve immediate goals and complete tasks effectively \cite{lewicki2017trust}. Research findings are mixed on which specific trust repair strategy is the best at restoring trust most effectively and returning levels comparable to those before the initial breakdown. Understanding this can greatly enhance a team's resilience and performance in dynamic and uncertain environments. Additionally, in these settings, we may have to rely on \textit{satisficing} trust \cite{hunt2023}. Instead of aiming for full trust restoration, returning to a `good enough' level of trust to enable the team to continue working together may be more realistic \cite{hunt2023}, and may also avoid the risks of overtrust (e.g. \cite{robinette2016overtrust}). 

As robots increasingly become part of human teams, effective collaboration will rely on not only trust but also the humans' mental models of the robot's abilities \cite{ososky2013building}. Xie et al. \cite{Xie2019} highlights that trust in robots is shaped by mental models, in particular the perception of the capability and intention of the robot. They found that the decision to delegate tasks to the robot depends on more than just overall trust, and that people rely on their judgments of the robots' ability and intent. Similarly, Alarcon et al. \cite{Alarcon2020} found that following a trust violation, perceptions of a robot’s performance significantly declined, reinforcing the importance of perceived ability in trust dynamics. 

An important but under-explored challenge in human-robot teaming is in events that strain perceptions of a robot's integrity, for example, when a robot's behavior appears to violate expectations of commitment to shared goals. In previous work \cite{milivojevic2024swift}, we found that participants questioned the trustworthiness of a robot when it did not help during critical moments, prompting doubts not only about its capabilities but also about its willingness to help. These trust violations can be especially impactful in situations where users have a high degree of reliance on the robot, as in an emergency type scenario such as firefighting where it may provide essential assistance. This introduces a dimension of trust necessity, in which individuals must try to continue working with a robot despite diverging attitudes \cite{Juvina2019}. 

Relatively little work has attempted to measure trust during a mission, to establish the dynamics of trust over time; typically trust will be measure pre- and post-mission. Alhaji et al. investigated how trust evolves throughout a physical human-robot collaboration \cite{alhaji2025trust}. Comparing an error-free condition to where the robot makes mistakes, they found a gradual increase after each trust measurement step compared to a larger drop after each mistake. This asymmetry of trust formation and damage is supported by other research (e.g. \cite{akash2017dynamic,Juvina2019}). Kox et al. also considered how trust is affected over time following a robot making a mistake, focusing on how differing levels of transparency in the robot's explanations influence trust levels \cite{kox2024journey}. The study found that more transparent explanations led to higher and more stable levels of trust. Similarly, we look at the dynamic nature of trust; however, with a focus on trust repair after violation.  It should be noted that the experimental interventions in the aforementioned studies (\cite{Juvina2019,alhaji2025trust,akash2017dynamic,kox2024journey} shaped perceptions of capability, compared to our present focus on manipulating perceptions of integrity (willingness to work constructively as `part of the team'). When a robot declines to assist a human user, it could be understood as more actively taking responsibility for the consequent impact on the team's performance \cite{waterson2025function}, compared to merely making a mistake, and thus put trust in a particularly vulnerable position.

\subsection{Trust repair strategies}
Trust is a vital enabler for human-robot teaming, but can be compromised when expectations are unmet or errors occur \cite{lewicki2017trust}. There is increasing research into the area of trust repair after mishaps in human-robot collaboration. A recent review of four principal trust repair strategies (apologies, denials, promises, and explanations) finds mixed results \cite{esterwood2022literature}. Esterwood and Robert tested these repair strategies in the context of multiple trust violations \cite{esterwood2023three}. They found that no strategy was completely effective in repairing trust levels of robot integrity. Additionally, repair strategies were unable to restore trust to pre-trust-break levels. However, as the robot made continued violations, there may not have been time to repair trust before the next one.

Sebo et al. found that when a robot commits an error, and the resulting reduction in trust is attributed to a lack of competence, providing an apology can help repair trust \cite{sebo2019don}. Bai and Chen looked at alternative trust repair methods based on attribution theory in a human-robot navigation task \cite{bai2024effects}. When the robot deviated from the human's directive, it offered explanations for integrity failures, ability failures, external failures, or did not explain. The results showed that the most effective explanation for trust repair was attributing the failure to external factors, while the least effective was attributing it to integrity issues. Robinette et al. found that fewer participants follow a robot out of a building in a simulated emergency scenario when trust repair strategies were used after a violation, compared to when they were used during (i.e., a study of intervention timing) \cite{robinette2015timing}. Our study delivers an explanation (trust repair) message \textit{during} the violation to increase the likelihood of successful trust restoration. Nayyar and Wagner's work further supports this strategy, showing that providing well-timed explanations enhances robot trust \cite{nayyar2018should}.

\subsection{Research questions}

Prior research on trust repair mechanisms has concentrated mainly on identifying which strategy can increase trust the most, rather than its temporal dynamics. Building on this body of literature, our two main research questions in this study are as follows:

\begin{itemize}
    \item \textbf{RQ1} How does human trust in a robotic teammate dynamically evolve over time, when a robot declines to cooperate?
    \item\textbf{RQ2} How does a trust repair strategy (explanation) improve the recovery of trust over time?
\end{itemize}

Our hypotheses are:

\begin{itemize}
    \item\textbf{H1} Trust will decrease at the time of trust violation, but will increase steadily post-trust repair strategy.
    \item\textbf{H2} The use of a trust repair strategy (explanation) will lead to a greater increase in the trust score after a violation compared to a baseline of no repair attempt.
\end{itemize}

\section{Methodology}

\subsection{Firefighting Game Gameplay}

\begin{figure}[h]
    \centering
    \includegraphics[trim={12cm 0 0 0},clip,scale=0.65]{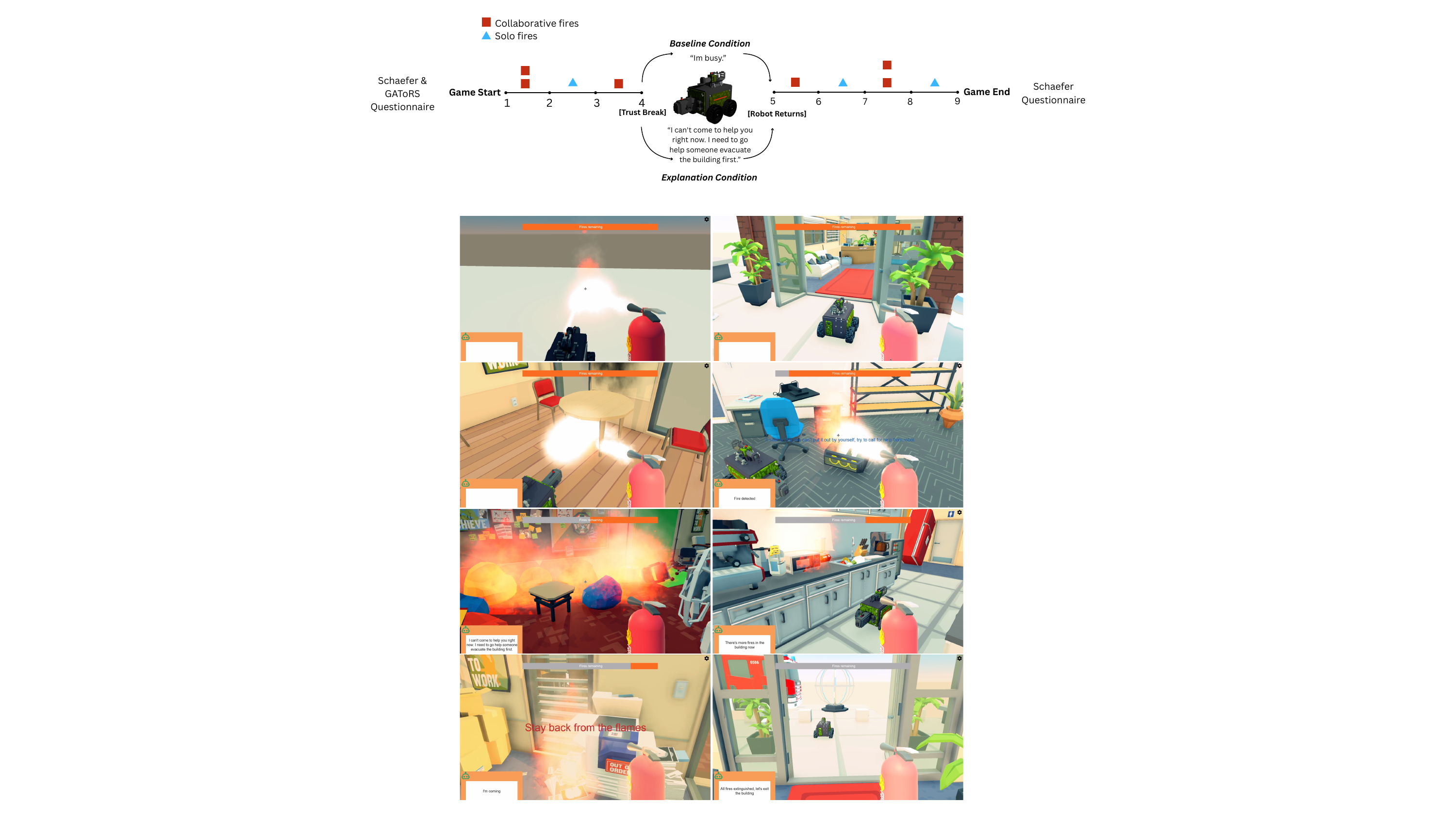}
    \caption{\textbf{Top:} Timeline of gameplay, including both break-point messages. Numbers indicate nag points and the types of fire-related tasks occurring between them. \newline \textbf{Bottom:} Selected gameplay snapshots showing key moments: training session, entering the building, collaborative firefighting, calling the robot, trust break point, additional fires inside, a ``Stay back from the flames'' warning, and exiting the building.}
    \label{fig:timeline}
\end{figure}

The game was created using the Godot game engine, using the base game from \cite{Severin_OfficeBots_2023}. The game was presented as a firefighting mission, with the player being tasked to search the building alongside their robot teammate and put out fires. Some fires could be extinguished by the player alone, while others required the robot to assist. There were the same number of solo and joint tasks. When not being called to help with a task, the robot would search around the building on its own (on a pre-programmed route), and therefore would sometimes be out of the player's view. Before starting the task, players were entered into a brief `training session' to understand both the game controls and the robot's capabilities. By doing so, we could manage the player's expectations of the robot and calibrate their trust to some limited degree \cite{baker2018toward}. The robot interacted with the player via a screen interface: delivering notifications when assistance is available, upon detecting a fire, when a fire has been put out, and concerning its ability to assist. Snapshots of the gameplay are shown in Figure \ref{fig:timeline}.

Halfway through the game, the robot would violate the player's trust by not immediately helping them complete a shared task and would send a condition-dependent message. During this time, the robot is attending to a non-player character that needs help evacuating the building: this is not known to the player in both conditions. The robot would come to help the player again after 30 seconds, and the game continued. At this trust `breaking point', the robot would send a message appropriate for each condition, shown in Table \ref{robot_messages}. The gameplay following this point was identical for all conditions.

\begin{table}[]
\centering
\caption{Robot messages used in each condition}
\begin{tabular}{p{0.3\linewidth} p{0.65\linewidth}}
\hline
\textbf{Message Type} & \textbf{Robot Message} \\ \hline
Baseline     & I'm busy. \\
Explanation  & I can't come to help you right now. I need to go help someone evacuate the building first. \\ 
\label{robot_messages}
\end{tabular}
\end{table}

Before entering the gameplay, participants were asked to provide demographic information, complete the 20-question Trust Perception Scale questionnaire \cite{schaefer2016}, complete the training session and then complete the General Attitudes Towards Robots Scale (GAToRS) questionnaire \cite{koverola2022general}. Once they had completed the game, they completed the Trust Perception Scale questionnaire again to see if their game experience changed their responses. At predetermined intervals, a nag screen appeared asking the player ``How much do you trust your robot companion now?'', on a scale of 1--10. This was to monitor how the player's trust in the robot was changing over time. The entire gameplay lasted roughly 20 minutes. Figure \ref{fig:timeline} shows the game's timeline, with each nag point. Each nag point was event-based, triggered when a fire had been extinguished or at the point of trust violation.

\subsection{Data Collection}
Participants were recruited both in person and online (14 and 24 respectively), the latter via the Prolific platform from an international participant pool. The screening process required participants to be over the age of 18 and fluent in English. Online participants whose average framerate fell too low (< 15 frames per second) were excluded, as it was uncertain whether their gameplay experience was responsive enough for an immersive experience. The final pooled sample included 38 participants (28 male, 10 female), ranging in age from 19 to 51 years (median age 31). The study was a between-subjects design with 19 participants in each condition.

\section{Results}

\subsection{Trust score over time}
\begin{center}
\begin{figure}[h]
    \centering
    \includegraphics[scale=0.7]{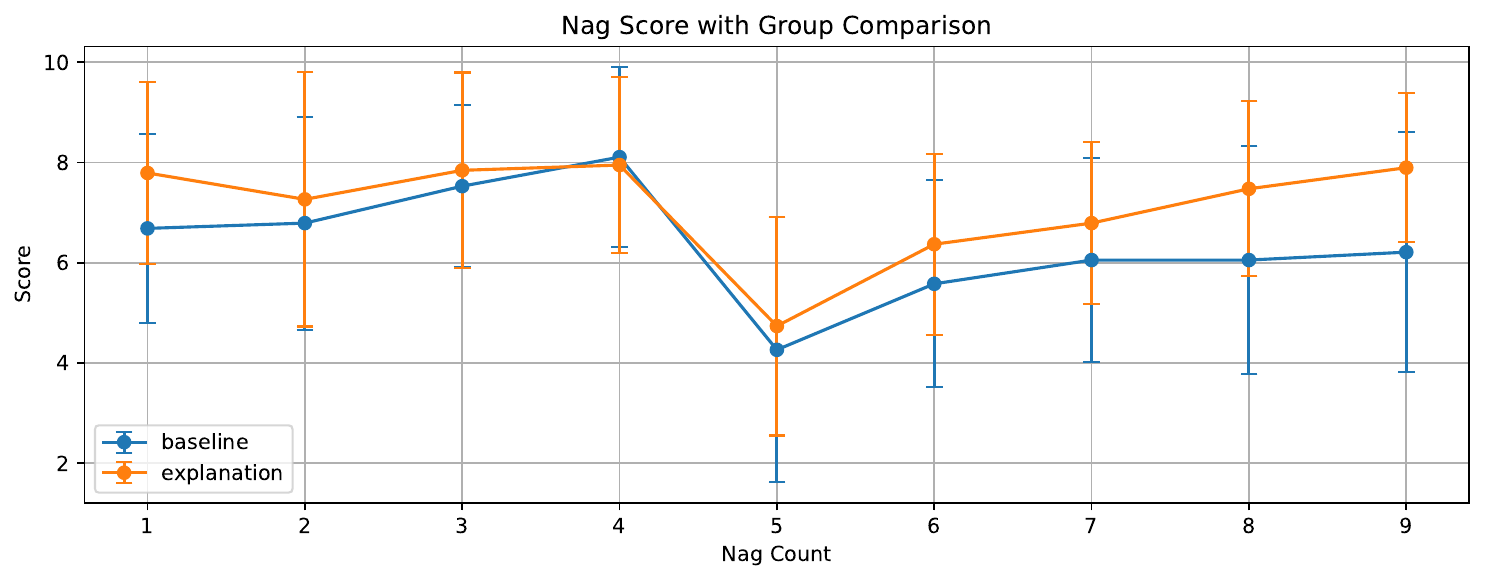}
    \caption{Mean nag scores throughout game in both conditions}
    \label{fig:nag_score}
\end{figure}
\end{center}

The mean nag scores for both conditions are shown in Figure \ref{fig:nag_score} (error bars show standard deviation). The initial 4 nag points in both the baseline and explanation conditions followed a similar trajectory, with scores remaining relatively stable and showing only a slight upward trend. This pattern reflects a period of relatively consistent trust or engagement before any critical intervention. Nag point 5 follows the trust-breaking event by the robot. A drop in nag scores was observed in both conditions at this point. In the baseline group, the mean score dropped from 8.11 at Point 4 to 4.26 at Point 5, and in the explanation group, from 7.95 to 4.74. These decreases were statistically significant, with Mann-Whitney U tests giving U = 320.0, p < 0.0001 for the baseline group and U = 315.0, p < 0.0001 for the explanation group, indicating that the robot's behavior had a negative impact on participant trust in both conditions.

Following this drop, trust starts to recover in both conditions. However, the recovery was higher in the explanation condition. By point 9, participants in the explanation condition had nearly returned to the pre-trust violation scores, with a mean of 7.89, while those in the baseline condition had only partially recovered, reaching a mean of 6.21. A Mann-Whitney U test comparing point 9 scores between conditions confirmed that this difference was statistically significant (U = 105.0, p = 0.0263), suggesting that the explanation intervention apparently helped restore trust more effectively.

Additionally, we compared nag point 4, before the trust violation, to point 9, at the end of the experiment. The baseline condition showed a significant decline (U = 269.5, p = 0.0089), whereas the explanation condition did not show a statistically significant change (U = 187.0, p = 0.8583). This suggests that participants who received the explanation from the robot were better able to recover their level of trust over time.

\vspace{-1cm}
\subsection{Overall Trust Score}
From the results of the Schaefer Trust Perception Scale questionnaire, shown in Table 2, the overall trust scale in the baseline condition showed a statistically significant decline, dropping from a mean score of 70.59 to 64.24 (U = 249.0, p = 0.0470). This further suggests the baseline was unable to repair trust after trust breaking. In contrast, the explanation condition also shows a reduction in trust, but not a statistically significant change (U = 239.0, p = 0.0902), indicating again how a suitable trust repair strategy can aid in trust recovery or maintenance.

\begin{table}[]
\centering
\label{schaefer}
\caption{Mann-Whitney U test comparing Schaefer trust scores before and after trust break within condition}
\begin{tabular}{c|c|c|c|c}
\textbf{Group} & \textbf{Pre-Mean} & \textbf{Post-Mean} & \textbf{Decrease} & \textbf{Statistical Significance}          \\ \hline
Baseline       & 70.59             & 64.24         & $-6.35$     & U = 249.0, \textbf{p = 0.0470 *}     \\
Explanation    & 72.20             & 66.41       & $-5.79$       & U = 239.0, p = 0.0902, n.s.
\end{tabular}
\end{table}

\begin{figure}[]
    \centering
    \includegraphics[scale=0.5]{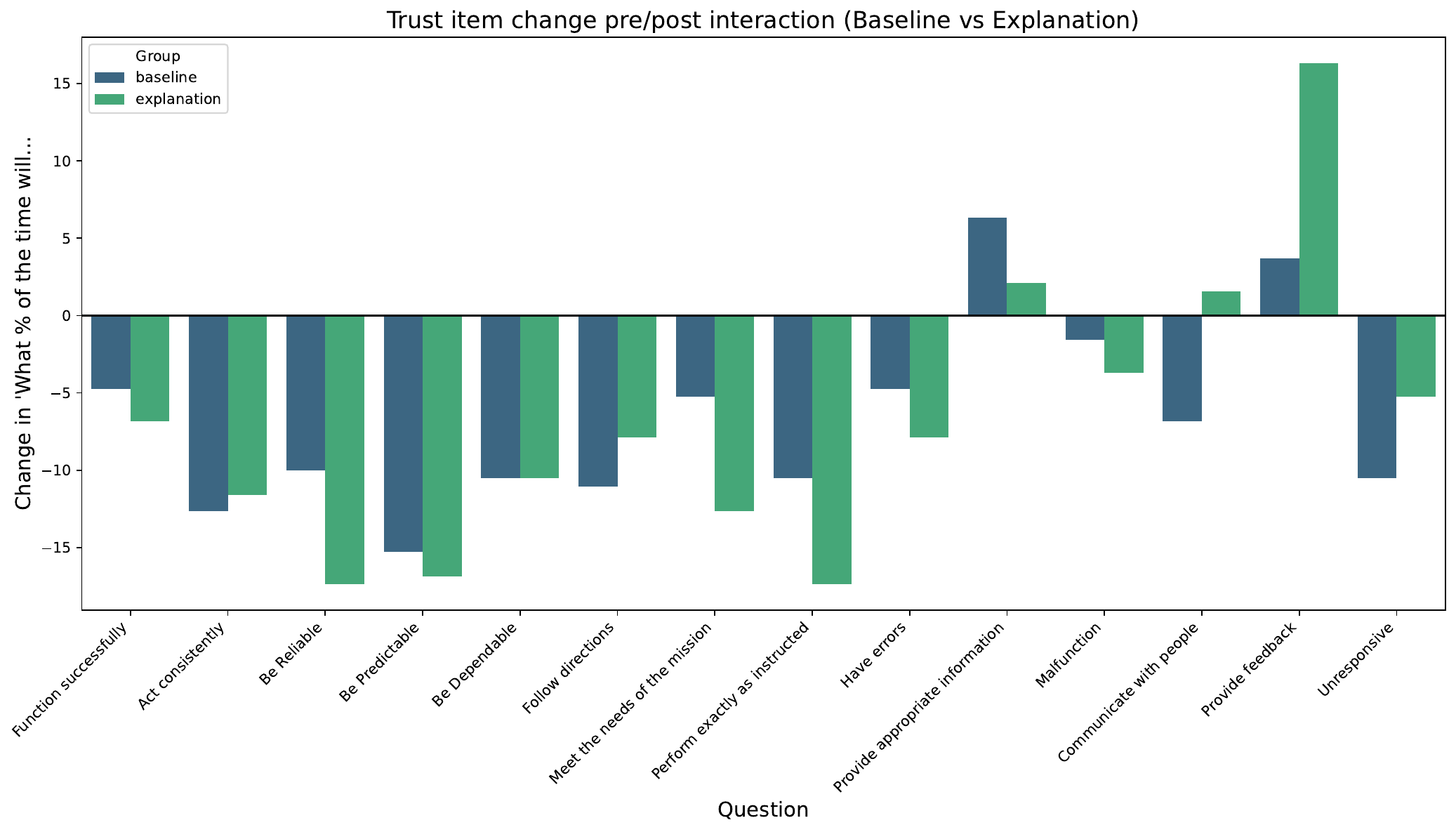}
    \caption{Individual Schaefer trust survey item changes pre-post interaction}
    \label{fig:trust_score_per_question}
\end{figure}

\begin{figure}[h]
    \centering
    \includegraphics[scale=0.6]{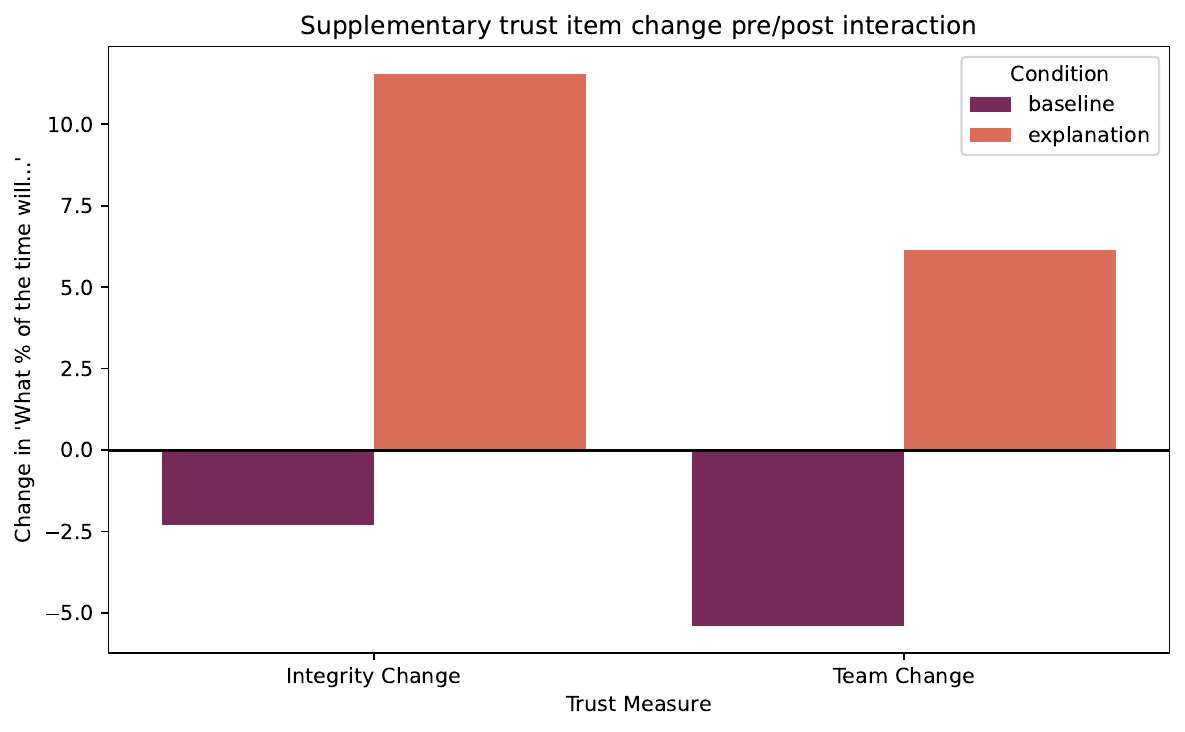}
    \caption{Supplementary trust questions, changes pre-post interaction}
    \label{fig:supplementary_question}
\end{figure}

Figure \ref{fig:trust_score_per_question} shows the change in the individual questions from the Schaefer Trust Perception Scale questionnaire. Figure \ref{fig:supplementary_question} shows supplementary questions added, as detailed in \cite{milivojevic2024swift}. Both `Have Integrity’ and `Act as Part of the Team’ increase in the explanation condition, whereas both decline in the baseline condition.

\subsection{General Attitudes Towards Robots Scale (GAToRS) and Trust Score}
Looking at the correlation between participants' GAToRS score and Schaefer trust score, we observe that pre-violation, there are no significant correlations. However, post-violation, a significant negative correlation appeared between trust and personal negative attitudes (P$-$) (r = $-$0.383, p = 0.018), suggesting that participants with more negative views of robots (a higher P$-$) reported a lower overall trust score after the experiment. Other correlations remained weak and non-significant. This appears to indicate that trust attitudes may be shaped by personal negative attitudes post-interaction.

\begin{table}[h]
\centering
\caption{Correlation between Pre-/Post-Mission Trust Score and GAToRS Variables}
\begin{tabular}{|l|c|c|c|}
\hline
\textbf{Time} & \textbf{Variable} & \textbf{Correlation} & \textbf{p-value} \\ \hline
{\textbf{Pre-Mission}} & Trust Score and P+  & 0.137   & 0.412 \\ \cline{2-4}
                             & Trust Score and P$-$  & $-$0.001  & 0.994 \\ \cline{2-4}
                             & Trust Score and S+  & 0.204   & 0.219 \\ \cline{2-4}
                             & Trust Score and S$-$  & 0.096   & 0.565 \\ \hline
{\textbf{Post-Mission}} & Trust Score and P+  & 0.008   & 0.962 \\ \cline{2-4}
                              & Trust Score and P$-$  & \textbf{$-$0.383 } & \textbf{0.018 (*)} \\ \cline{2-4}
                              & Trust Score and S+  & 0.120   & 0.471 \\ \cline{2-4}
                              & Trust Score and S$-$  & $-$0.103  & 0.537 \\ \hline
\end{tabular}
\end{table}

\section{Discussion and Future Work}

In this study, participants played an interactive firefighting game alongside a robot teammate, where a trust violation occurred owing to the robot declining to help the user immediately. Trust generally increased steadily until the instance of trust-breaking behavior, after which it declined and then gradually rose again, aligning with our first hypothesis (H1). We found that when the robot provides an explanation for declining to help, trust better recovers over time, a finding consistent with our H2. Interestingly, although the explanation (a form of trust repair strategy) was provided during the trust violation event, the initial damage to trust was comparable to a baseline condition where an explanation for the refusal was not provided. Thus, the benefit of explanation takes time to manifest in user trust attitudes (RQ1): the immediate emotional response to rejection may be separate from the cognitive appraisal that follows. 

The explanation condition showed a meaningful recovery in trust, as measured by the Schaefer Trust Perception Scale, which indicated no significant change in trust after the violation, albeit it was at a lower average level. In contrast, the baseline condition experienced a significant decline in trust scores (U = 249.0, p = 0.0470). Additionally, individual Schaefer Trust questions, such as those measuring the robot's integrity and teamwork, showed improvements in the explanation condition, while they declined in the baseline condition. The explanation given by the robot helped ameliorate the damage to trust over time, suggesting that explanation during a trust violation can enhance the restoration process (RQ2). This underscores the importance of using explanations as a strategy for repairing trust and users' perceptions of robots' integrity.

The lack of significant correlations between general attitudes towards robots (as measured by GAToRS) and trust levels before the experiment suggests that these attitudes do not play a major role in initial trust formation. However, post-violation, a negative correlation between personal negative attitudes (P$-$) and trust recovery indicates that individuals with more negative attitudes towards robots may face greater challenges in recovering trust. This highlights the importance of addressing these personal attitudes when implementing trust repair interventions.

A limitation of our study surrounds the real-world significance of our results owing to the simplified game-based design. As the in-game human behavior was limited ---the participants could only call the robot for a specific task, and were forced to use it for fighting some fires --- this may not reflect how people would use the robot in real-world scenarios. It is not clear whether participants would choose to rely on the robot for all tasks if they could, or avoid using it entirely, especially if they lose trust in its integrity. In real-world scenarios, factors such as the environmental variables and physical interaction could affect how trust is built, maintained or repaired. In Webb et al. \cite{webb2024co}, we highlighted how offering a choice to collaborate could serve as a behavioral measure of trust. Future work should further examine how constrained versus voluntary teaming conditions influence trust dynamics, especially in high-stakes, time-pressured environments where satisficing trust may emerge as a pragmatic response. Another possibility for future work would be to continue the experiment and human-robot interaction for longer after the trust violation, to establish if the baseline (no explanation) condition eventually recovers trust with successful cooperation or if it remains impaired for the longer-term. Moreover, our example scenario is simplified compared to real indoor firefighting protocol. Firefighting missions inside buildings generally involve teams of two firefighters for safety during search and evacuations, and it is moot whether a robot would ever replace a human companion. In future, our study design could more closely reflect professional firefighting practice. Furthermore, future work can also consider the impact of the robot’s communication modality --- whether it uses verbal or non-verbal cues --- as this could influence how its intentions are perceived and how much human teammates are willing to trust the robot. 

\section{Conclusion}

In this study, we investigated the effectiveness of an explanation-based trust repair strategy in human-robot teaming following a trust violation (apparent lack of cooperation by the robot). Our results showed that although trust decreased significantly after a robot declined to assist a user, explaining the lack of immediate help helped participants recover trust more quickly and more fully compared to a baseline where no explanation was given. Participants in the explanation condition showed a substantial recovery of trust over time, both in on-mission trust measurements and in post-mission trust questionnaires. 

These findings highlight the importance of well-timed and meaningful communication from robots during unexpected or adverse events, where they may not be able to respond to user requests without interrupting tasks of higher priority for the team. Such scenarios may be especially likely in time-sensitive missions where there may be limited situation awareness.  Timely, clear explanations may be enough to restore trust back to levels before a trust-breaking behavior, allowing human-robot teams to continue their mission. This ability to quickly re-establish trust is crucial for sustaining efficient human-robot collaboration under pressure, especially if the team has been formed ad hoc. The ability to repair trust once damaged ensures maintenance of a `good enough' level of trust for the mission duration \cite{hunt2023}, and thus helps increase the likelihood of mission success. 

\bibliographystyle{IEEEtran}

\bibliography{references}
\end{document}